# Fractional Quantum Hall Effect States on a Torus show a Liquid – Insulator Second Order Transition.

J. Montes

jmn@usal.es

Departamento de Física Aplicada, Universidad de Salamanca,

Patio de Escuelas 1, 37008, Salamanca. Spain.

## Abstract

From the Fractional Quantum Hall Effect States on a torus with filling factor $\nu = 1/p$, p odd, we found that for small values of $\nu$ such states describe triangular Wigner crystals which are stable configurations. Therefore, the insulator phase of the two-dimensional electron system can be described with the same wave function that the liquid phase, and the incompressible liquid freezes (or the Wigner crystal melts) without any quantum transition between states, thus the system phase changes by means of a continous (second-order) transition. We found at least one stable configuration of a Wigner crystal with F = 3 electrons for cell and we investigate the behavior of the crystal when the lengths ratio $L_y/L_x$ changes.

# 1. Introduction

The Fractional Quantum Hall Effect (FQHE) is theoretically understood in terms of the Laughlin proposal [1] of a many-body variational ground state of Jastrow type. This approach works very well for odd inverse filling factors; there is a hierarchy labelled by the integer ratio between the number of states in the first Landau level and the number of electrons. For the first few members of the hierarchy Laughlin wave functions describe an incompressible quantum fluid but it was very soon conjectured that a "phase transition" to a Wigner crystal of equally spaced electrons would occur for lower densities and higher magnetic fields [2] and in fact triangular arrays of this kind have been observed in the FQHE [3]. The key observation of our article is the interpretation of the square of the ground state wave function as the probability distribution of a classical statistical system: a 2-dimensional Coulomb plasma of charged particles moving in a constant charge density "neutralizing" background. The physical parameters, effective temperature, particle charge and background charge density are related to the inverse of the filling factor in such a way that the statements above follow from general principles of statistical mechanics.

Laughlin-Jastrow wave functions for the FQHE on a torus has been proposed by Haldane and Rezayi [4] in terms of Riemann Theta functions. Our first goal is therefore to identify the 2-dimensional classical plasma described by the FQHE ground state. A priori the search for equilibrium distributions of such a system seems a formidable task. Owing to the periodicity properties of the potential we will show how indeed a triangular Wigner crystal arises as an equilibrium distribution by using an *ansatz* adapted to the periodicity properties. Our solutions are in good agreement with the vortex distribution of a fluid on a torus found in [5] in a mathematical analogous problem.

## 2. Plasma Potential of FQHE States on a Torus.

We consider a periodic lattice in the complex plane having as unit cell the domain bounded by the four points: $z = \{0, 1, \tau, \tau + 1\}$, where $\tau$, a complex number such that $\mathrm{Im}\,\tau > 0$, determines de axis and length of the second period assuming that the first period is on the OX axis direction and choosing units where $L_x = 1$. We are concerned with the motion of a two-dimensional electron system (2DES) in this periodic lattice under a perpendicular magnetic field with constant value $B = \dfrac{k\Phi_0}{A}$ where k is an integer; $\Phi_0 = \dfrac{hc}{e}$ is the flux quantum and $A = \mathrm{Im}\,\tau$ is the area of the principal cell. If N is the total number of electrons, then the filling factor is $\nu = \dfrac{1}{p} = \dfrac{N}{k}$ and shows that the number k of flux quanta crossing the unit cell is equal to the degeneration of Landau levels. Assuming $p > 1$ and odd, Haldane and Rezayi in [4] proposed for the ground state of a 2DES with periodic boundary conditions in the *plateau* 1/p the following wave function of the Laughlin-Jastrow type:

$$\Psi_m^s(z_1,...,z_N) = \Theta\!\begin{bmatrix} s/2 \\ s/2 + m/p \end{bmatrix}\!(Z|\tau/p)\exp\!\left\{-\dfrac{\mathrm{Im}\,\tau}{2 l_B^2}\sum_{i=1}^{N}(\mathrm{Im}\,z_i)^2\right\}\prod_{i<j}\!\left(\Theta\!\begin{bmatrix} 1/2 \\ 1/2 \end{bmatrix}\!(z_i - z_j|\tau)\right)^{p}$$

$$s = 0, 1\ ;\ m = 0, 1,\ldots, p\text{-}1 \qquad (2.1)$$

This collective wave function has good features: it approximates well enough the first eigenstate of the many-body Schrödinger equation with periodic conditions; it exhibits the right occupation of the Landau levels; it verifies the Fermi statistics and therefore it is accepted that provides the basic explanation of the FQHE in a periodic lattice.

In formula (2.1) we have taken a Landau gauge for the electromagnetic potential vector, whereas $l_B = \sqrt{\frac{\mathrm{Im}\,\tau}{2\pi k}}$ is the magnetic length (in units where $\Phi_0 = 1$); $Z = \sum_{i=1}^{N} z_i$ is the center of mass complex coordinate and $\Theta\begin{bmatrix} a \\ b \end{bmatrix}(z|\tau)$ are the Theta Functions of rational characteristic, see [6], [7]. There is a novelty with respect to the non-periodic case because a degeneration due to the center of mass factor of the wave function arises. We must choose s = 0 if N is odd and s = 1 for N even, but we are free to fix $m \in \mathbf{Z}_p$. Note that we have chosen a center of mass wave function different from which Haldane and Rezayi adopt in [4], but this function satisfies the boundary conditions also well.

The quantum probability density $|\Psi_0^s|^2$ can be interpreted as a classical Boltzman distribution according to:

$$|\Psi_0^s|^2 = e^{-\beta V_s},$$

where $\beta = 2\pi k/\alpha$, given $\alpha = L_y/L_x = |\tau|$, is the inverse of the effective temperature and the plasma potential $V_s$ is given by the formula:

$$V_s = \sum_{i=1}^{N}(\mathrm{Im}\,z_i)^2 - \frac{\alpha}{\pi p N}\log\left|\Theta\begin{bmatrix} s/2 \\ s/2 \end{bmatrix}(Z|\tau/p)\right| - \frac{\alpha}{2\pi N}\sum_{i,j}\log\left|\Theta\begin{bmatrix} 1/2 \\ 1/2 \end{bmatrix}(z_i - z_j|\tau)\right| \quad (2.2)$$

The minima of this potential, or equivalently the maxima of probability density, will be investigated next.

# 3. Stationary Probability Distributions.

Extremal configurations satisfy the equations, for each i = 1, 2,..., N:

$$\operatorname{Re} G_i = 0 \quad ; \quad \operatorname{Im} G_i = -\frac{2\pi p N}{\alpha} \operatorname{Im} z_i$$

$$G_i = \frac{\Theta'\begin{bmatrix} s/2 \\ s/2 \end{bmatrix}}{\Theta \begin{bmatrix} s/2 \\ s/2 \end{bmatrix}}(Z|\tau/p) + p \sum_{j \neq i} \frac{\Theta'\begin{bmatrix} 1/2 \\ 1/2 \end{bmatrix}}{\Theta \begin{bmatrix} 1/2 \\ 1/2 \end{bmatrix}}(z_i - z_j|\tau) \qquad (3.1)$$

Equations (3.1) seem difficult but fortunately they have a great deal of symmetry. To unveil the symmetry and rewrite (3.1) in a simpler manner, we make the following *ansatz* in order to look for crystalline solutions:

$$z_{h+Fa} = Z_h + w_a \quad , \quad h = 0, 1,..., N_c - 1 \; ; \; a = 1, 2,..., F \;.$$

Here, $Z_h$ is the complex coordinate of the center of the h cell, $w_a$ is the relative coordinate of the particle i = h + Fa with respect to $Z_h$, F is the number of electrons per cell and $N_c$ the number of cells: $N = N_c F$. From the periodicity conditions we deduce:

$$Z_h = n_h + m_h \tau \quad ; \quad n_h, m_h \in \mathbf{Z}$$

$$Z = \sum_{h=0}^{N_c-1} \sum_{a=1}^{F} (Z_h + w_a) = F \sum_{h=0}^{N_c-1} n_h + N_c w \qquad (3.2)$$

where w are the center of mass coordinates of the particles in the principal cell, and provided that the axis are chosen such that $\sum_{h=0}^{N_c-1} m_h = 0$ (symmetry with respect to OX axis of the whole crystal). In addition, it is verified:

1.
$$\left|\Theta\begin{bmatrix}s/2\\s/2\end{bmatrix}(Z|\tau/p)\right| = \left|\Theta\begin{bmatrix}s/2\\s/2\end{bmatrix}(N_c w|\tau/p)\right| \qquad (3.3\text{ a})$$

This follows from the invariance of the modulus of the Theta functions under integer translations.

2.
$$\sum_{i=1}^{N}(\operatorname{Im} z_i)^2 = N_c \sum_{a=1}^{F}(\operatorname{Im} w_a)^2 + const. \qquad (3.3\text{ b})$$

3.
$$\sum_{h,k}\sum_{a<b}\log\left|\Theta\begin{bmatrix}1/2\\1/2\end{bmatrix}(Z_h - Z_k + w_a - w_b|\tau)\right|^2 =$$
$$= N_c^2 \sum_{a<b}\log\left|\Theta\begin{bmatrix}1/2\\1/2\end{bmatrix}(w_a - w_b|\tau)\right|^2 + const. \qquad (3.3\text{ c})$$

This comes from the equation:

$$\Theta\begin{bmatrix}1/2\\1/2\end{bmatrix}(z + n + m\tau|\tau) = (-1)^{n+m}\exp\left[-2\pi i\left(mz + \frac{1}{2}m^2\tau\right)\right]\Theta\begin{bmatrix}1/2\\1/2\end{bmatrix}(z|\tau)$$

The result is that using (3.3) we can write the following formula, not very different from (2.2):

$$V_s = N_c \sum_{a=1}^{F} (\mathrm{Im}\, w_a)^2 - \frac{\alpha}{\pi p N} \log \left| \Theta \begin{bmatrix} s/2 \\ s/2 \end{bmatrix} (N_c w | \tau / p) \right| - \frac{\alpha N_c}{2\pi F} \sum_{a,b} \log \left| \Theta \begin{bmatrix} 1/2 \\ 1/2 \end{bmatrix} (w_a - w_b | \tau) \right| \quad (3.4)$$

The stationary configurations satisfy:

$$\mathrm{Re}\, H_a = 0 \;\; ; \;\; \mathrm{Im}\, H_a = -\frac{2\pi p N}{\alpha} \mathrm{Im}\, w_a$$

$$H_a = H^{CM}(w) + H_a^R(w_b) =$$
$$= \frac{\Theta' \begin{bmatrix} s/2 \\ s/2 \end{bmatrix}}{\Theta \begin{bmatrix} s/2 \\ s/2 \end{bmatrix}} (N_c w | \tau / p) + p N_c \sum_{b \neq a} \frac{\Theta' \begin{bmatrix} 1/2 \\ 1/2 \end{bmatrix}}{\Theta \begin{bmatrix} 1/2 \\ 1/2 \end{bmatrix}} (w_a - w_b | \tau) \quad (3.5)$$

*i.e.* we obtain the same equations as in (3.1) but only referred to the principal cell coordinates and with the renormalization of p to $pN_c$. Because it is verified $\sum_a H_a^R = 0$, it is possible to separate the center of mass coordinates, which satisfy the equations:

$$\mathrm{Re}\, H^{CM} = 0 \;\; ; \;\; \mathrm{Im}\, H^{CM} = -\frac{\pi p N_c}{\alpha} \mathrm{Im}\, w \;.$$

This subsystem admits the solution w = n + ½ for N even, or w = n for N odd, where n is an arbitrary integer. We are left with the following equations which only depends on the characteristics of one cell:

$$\mathrm{Re}\, H_a^R = 0 \;\; ; \;\; \mathrm{Im}\, H_a^R = -\frac{2\pi F}{\alpha} \mathrm{Im}\, w_a \;.$$

Their solutions, if stable, form a crystal for large p because the electrons located at the equilibrium points in the lattice dominate the probability distribution. Numerical studies for a small number of electrons (N ≤ 17) suggest that the transition between the incompressible liquid and the Wigner crystal happens for p ≥ 9 [8].

The stability of the stationary configurations is read from the spectrum of the Hessian. If we define:

$$K_s(z|\tau) = \frac{d^2}{dz^2} \log \Theta \begin{bmatrix} s/2 \\ s/2 \end{bmatrix}(z|\tau) \ ;$$

$$K_0(w) = K_s(N_c w | \tau/p) \ ;$$

$$K_{a,b} = K_1(w_a - w_b | \tau) \ ;$$

if in order to be specific, we take $\sum_a w_a = n \in \mathbf{Z}$ (N odd); if we introduce the notations $x_i = \operatorname{Re} z_i$, $y_i = \operatorname{Im} z_i$ with $i = h + Fa$, $j = k + Fb$; then the Hessian of the plasma potential over a stationary configuration can be computed from the following formulae:

$$\begin{aligned}
\frac{\partial^2 V_s}{\partial x_i \partial x_j} &= -\frac{\alpha}{\pi p F}\left(\operatorname{Re} K_0(n) - p \operatorname{Re} K_{a,b}\right) && i \neq j \\
&= -\frac{\alpha}{\pi p F}\left(\operatorname{Re} K_0(n) + p N_c \sum_c \operatorname{Re} K_{a,c}\right) && i = j
\end{aligned} \quad (3.6 \text{ a})$$

$$\begin{aligned}
\frac{\partial^2 V_s}{\partial x_i \partial y_j} &= \frac{\alpha}{\pi p F}\left(\operatorname{Im} K_0(n) - p \operatorname{Im} K_{a,b}\right) && i \neq j \\
&= \frac{\alpha}{\pi p F}\left(\operatorname{Im} K_0(n) + p N_c \sum_{c \neq a} \operatorname{Im} K_{a,c}\right) && i = j
\end{aligned} \quad (3.6 \text{ b})$$

$$\frac{\partial^2 V_s}{\partial x_i \partial x_j} = -\frac{\partial^2 V_s}{\partial y_i \partial y_j} + 2 N_c \delta_{i,j} \quad (3.6 \text{ c})$$

Inspired in (3.6 c) we define the plasma potential for cell $W_s = V_s / N_c$ and we take the limit $N_c \to \infty$. With this limit we are taking in fact the thermodynamic limit because both the number of electrons (= $N_c F$) and the total area of the sample (= $N_c \, \text{Im} \, \tau$) tend to infinity, whereas the density (= $F/\text{Im} \, \tau$) remains finite. The result for the derivatives is:

$$\frac{\partial^2 W_s}{\partial x_i \partial x_j} = 0 \qquad i \neq j$$
$$= -\frac{\alpha}{\pi F} \sum_c \text{Re} K_{a,c} \qquad i = j \qquad (3.7 \text{ a})$$

$$\frac{\partial^2 W_s}{\partial x_i \partial y_j} = 0 \qquad i \neq j$$
$$= \frac{\alpha}{\pi F} \sum_c \text{Im} K_{a,c} \qquad i = j \qquad (3.7 \text{ b})$$

$$\frac{\partial^2 W_s}{\partial x_i \partial x_j} = -\frac{\partial^2 W_s}{\partial y_i \partial y_j} + 2\delta_{i,j} \qquad (3.7 \text{ c})$$

Now, the Hessian of the potential for cell only depends on the electron relative coordinates in the principal cell and the problem of analyzing their spectrum reduces also to a problem in only one cell. It is important to notice that the minima of the potential are independent of the filling factor 1/p just like their Hessian. Nevertheless we will not consider that these results are valid for low values of p, since in this case nobody have found in any 2DES a Wigner crystal but a fluid behavior. We will suppose that suitable FQHE states which can approximate Wigner crystals do not occur except for great values of p ($\geq 9$).

## 4. Wigner Crystals.

The variational problem (3.4) - (3.7) is, in spite of the simplifications, too difficult for an analytical treatment. We offer some numerical analysis.

A) Choosing $\tau = I$ (square lattice) and $F = 3$ we find a solution with three nodes of the potential at the positions (see Fig.1):

| node | x | y |
|------|------|---------|
| 1 | 0.5 | 0.27614 |
| 2 | 0.5 | -0.27614 |
| 3 | -1 | 0 |

This periodic solution is indeed an equilibrium configuration forming a triangular Wigner crystal: the eigenvalues and eigenvectors with respect to the coordinates $(x_1, x_2, x_3, y_1, y_2, y_3)$ are:

$$\lambda_1 = 0.15469 \quad \mathbf{u}_1 = (1,0,0,0,0,0)$$
$$\lambda_2 = 0.15469 \quad \mathbf{u}_2 = (0,1,0,0,0,0)$$
$$\lambda_3 = 1,8453 \quad \mathbf{u}_3 = (0,0,0,1,0,0)$$
$$\lambda_4 = 1,8453 \quad \mathbf{u}_4 = (0,0,0,0,1,0)$$
$$\lambda_5 = 1.1576 \quad \mathbf{u}_5 = (0,0,1,0,0,0)$$
$$\lambda_6 = 0.84243 \quad \mathbf{u}_6 = (0,0,0,0,0,1)$$

and it verifies $\lambda_1 + \lambda_3 = \lambda_2 + \lambda_4 = \lambda_5 + \lambda_6 = 2$. The eigenvectors are simple translations of the nodes in the directions of the axis.

We have also analysed how the previous results are modified when $\alpha = |\tau|$ varies in the closed interval $[0.5, 3]$, obtaining:

- The nodal positions only change slightly. There is a small increase in the width of the longitudinal bands of electrons in Fig. 1 at the same time that the bands tends to separate to each other when $|\tau|$ increases. The symmetry of the principal cell suffers no change.
- The eigenvectors are independent of $|\tau|$: the basis in which the Hessian is diagonal is insensitive to variations in $|\tau|$.
- For some value of lengths ratio such that $0.5 < |\tau| < 1$, the eigenvalue $\lambda_1$ changes the sign and the stationary configuration becomes unstable against small oscillations in the $\mathbf{u}_1$ mode, according to the numerical simulation shown in Fig. 2. Bellow this critical value of $|\tau|$ the crystal melts. All the other eigenvalues remain positive.
- The eigenvalues tend to asymptotic values when $|\tau| \to \infty$.

B) According to the results of Fukuyama *et al* [9] based on the Hartree-Fock approach for the 2DES in the FQHE, one might expect that by increasing F stronger instabilities would arise and this seems indeed to be the case. We present an stationary solution found by numerical computation for $\tau = 2I$ and $F = 4$:

| node | x | y |
|---|---|---|
| 1 | -1.5 | 0.36778 |
| 2 | 0 | -0.36778 |
| 3 | -0.75 | 0 |
| 4 | 2.75 | 0 |

The eigenvalues and eigenvectors of the Hessian are:

$$\lambda_1 = -0.581 \quad \mathbf{u}_1 = (0.36, 0.36, -0.36, -0.36, 0, 0, 0.49, -0.49)$$

$$\lambda_2 = 0.184 \quad \mathbf{u}_2 = \frac{1}{\sqrt{2}}(-1, 1, 0, 0, 0, 0, 0, 0)$$

$$\lambda_3 = 0.828 \quad \mathbf{u}_3 = \frac{1}{2}(0, 0, 0, 0, 1, 1, 1, 1)$$

$$\lambda_4 = 0.869 \quad \mathbf{u}_4 = (0, 0, 0.51, -0.51, 0.35, 0.35, -0.35, -0.35)$$

$$\lambda_5 = 1.13 \quad \mathbf{u}_5 = (-0.35, -0.35, 0.35, 0.35, 0, 0, 0.51, -0.51)$$

$$\lambda_6 = 1.17 \quad \mathbf{u}_6 = \frac{1}{2}(1, 1, 1, 1, 0, 0, 0, 0)$$

$$\lambda_7 = 1.82 \quad \mathbf{u}_7 = \frac{1}{\sqrt{2}}(0, 0, 0, 0, -1, 1, 0, 0)$$

$$\lambda_8 = 2.58 \quad \mathbf{u}_8 = (0, 0, -0.49, 0.49, 0.36, 0.36, -0.36, -0.36)$$

The novelty is that $\lambda_1(|\tau|) < 0$ for any value of $|\tau|$, even as $|\tau| \to \infty$, and the crystal melts. The same characteristics appear for the numerical solution in the F = 5 and all $|\tau|$ case so that the expectations based on Fukuyama studies are confirmed.

## 5. Discussion.

We have found a stable equilibrium of the plasma potential with 3 electrons for cell and the solution has a great deal of simplicity because there are some exact numbers in the coordinates of the nodes; fixed relations between the eigenvalues of the Hessian, and slightly modifications of the crystal properties with the lengths ratio. The stability of the stationary configurations is a rich property of the FQHE states on a torus and is not always assured. An open question remains because it can have more stable equilibria and other type of Wigner crystals can arise.

## 6. Acknowledgements

I especially thank to Juan Mateos Guilarte for many aids and discussions in the writing of this article.

**References.**

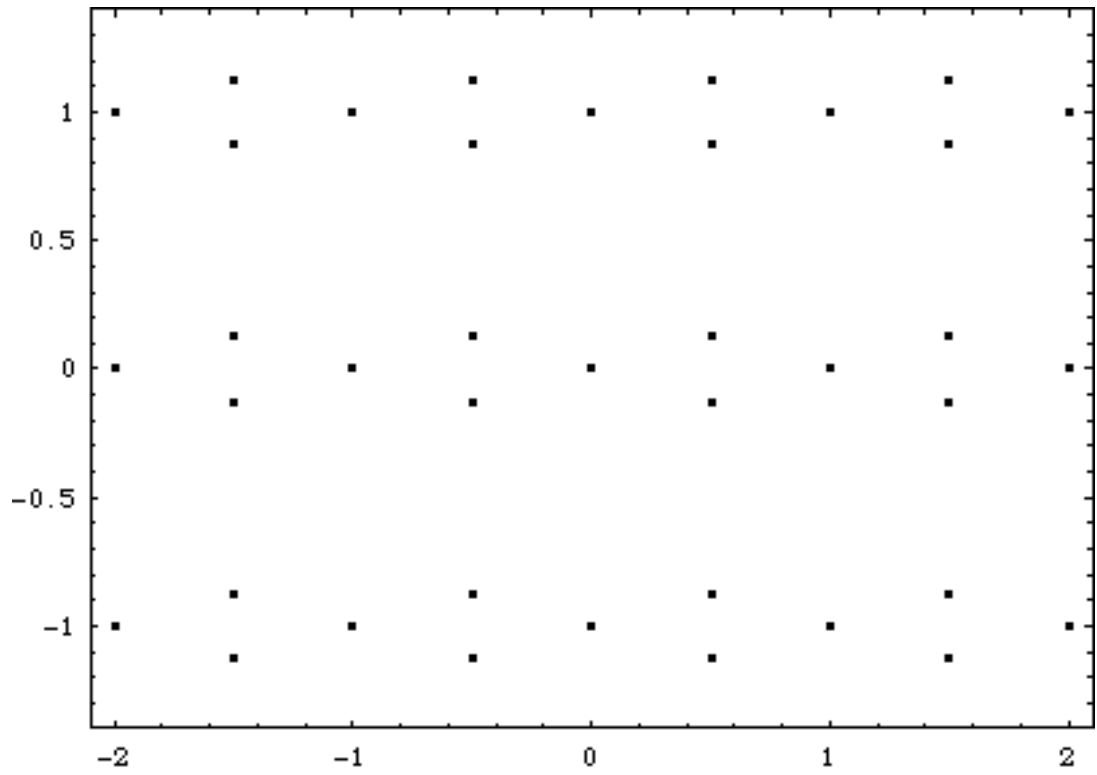

The Wigner crystal with 3 electrons for cell

Figure 1

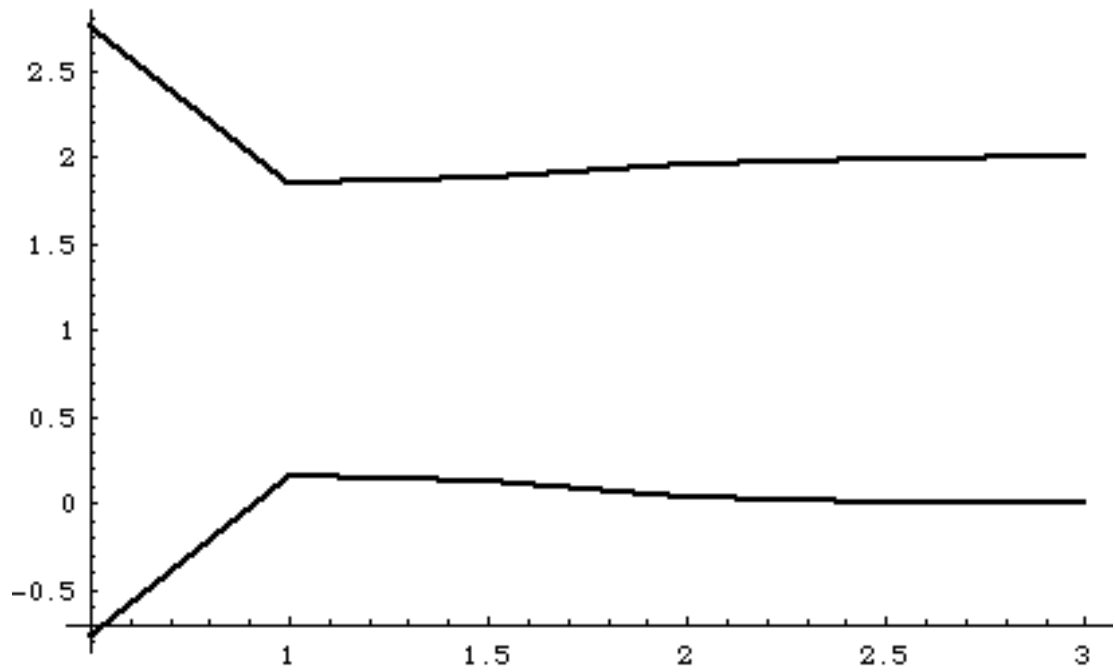

Plot of $\lambda_1$ (bottom) and $\lambda_3$ (top) versus $|\tau|$

Figure 2